\documentclass[conference]{IEEEtran}
\IEEEoverridecommandlockouts

\usepackage{cite}
\usepackage{amsmath,amssymb,amsfonts}
\usepackage{algorithmic}
\usepackage{graphicx}
\usepackage{textcomp}
\usepackage{xcolor}
\usepackage{tikz}
\usepackage[hyphens]{url} 
\usepackage{hyperref}     
\hypersetup{
    breaklinks=true,      
}
\usepackage{booktabs}
\def\BibTeX{{\rm B\kern-.05em{\sc i\kern-.025em b}\kern-.08em
    T\kern-.1667em\lower.7ex\hbox{E}\kern-.125emX}}
\begin{document}

\title{PIFS-Rec: Process-In-Fabric-Switch for Large-Scale Recommendation System Inferences\\
}

\author{\IEEEauthorblockN{Pingyi Huo}
\IEEEauthorblockA{\textit{The Pennsylvania State University} \\
University Park, PA, USA \\
pqh5140@psu.edu}
\and
\IEEEauthorblockN{Anusha Devulapally}
\IEEEauthorblockA{\textit{The Pennsylvania State University} \\
University Park, PA, USA \\
akd5994@psu.edu}
\and
\IEEEauthorblockN{Hasan Al Maruf}
\IEEEauthorblockA{\textit{AMD, Inc} \\
Austin, TX, USA \\
hasan.maruf@amd.com}
\and
\IEEEauthorblockN{Minseo Park}
\IEEEauthorblockA{\textit{AMD, Inc} \\
Austin, TX, USA \\
minseo.park@amd.com}
\and
\IEEEauthorblockN{Krishnakumar Nair}
\IEEEauthorblockA{\textit{AMD, Inc} \\
Austin, TX, USA \\
krishnakumar.nair@amd.com}
\and
\IEEEauthorblockN{Meena Arunachalam}
\IEEEauthorblockA{\textit{AMD, Inc} \\
Austin, TX, USA \\
meena.arunachalam@amd.com}
\and
\IEEEauthorblockN{Gulsum Gudukbay Akbulut}
\IEEEauthorblockA{\textit{The Pennsylvania State University} \\
University Park, PA, USA \\
gulsum@psu.edu}
\and
\IEEEauthorblockN{Mahmut Taylan Kandemir}
\IEEEauthorblockA{\textit{The Pennsylvania State University} \\
University Park, PA, USA \\
mtk2@psu.edu}
\and
\IEEEauthorblockN{Vijaykrishnan Narayanan}
\IEEEauthorblockA{\textit{The Pennsylvania State University} \\
University Park, PA, USA \\
vxn9@psu.edu}
}

\maketitle

\begin{abstract}
Deep Learning Recommendation Models (DLRMs) have become increasingly popular and prevalent in today's datacenters, consuming most of the AI inference cycles. The performance of DLRMs is heavily influenced by available bandwidth due to their large vector sizes in embedding tables and concurrent accesses. To achieve substantial improvements over existing solutions, novel approaches towards DLRM optimization are needed, especially, in the context of emerging interconnect technologies like CXL. This study delves into exploring CXL-enabled systems, implementing a process-in-fabric-switch (PIFS) solution to accelerate DLRMs while optimizing their memory and bandwidth scalability. We present an in-depth characterization of industry-scale DLRM workloads running on CXL-ready systems, identifying the predominant bottlenecks in existing CXL systems. We, therefore, propose PIFS-Rec, a PIFS-based scheme that implements near-data processing through downstream ports of the fabric switch. PIFS-Rec achieves a latency that is 3.89$\times$ lower than Pond, an industry-standard CXL-based system, and also outperforms BEACON, a state-of-the-art scheme, by 2.03$\times$.
\end{abstract}


\section{Introduction}

Personalized recommendation systems have emerged as a cornerstone in the interface between users and technology, spanning various application domains from e-commerce to social networking~\cite{naumov2019deep}. These systems, powered by deep learning techniques, sift through vast user data to deliver tailored content. This personalized approach boosts user engagement and significantly enhances overall satisfaction. As these systems become increasingly essential components of our digital experiences, datacenters worldwide are scaling up their capabilities, dedicating extensive resources to the AI inference tasks that underpin recommendation models. 

Among various deployed models~\cite{carbonell1983overview, jordan2015machine,baltruvsaitis2018multimodal}, Deep Learning Recommendation Models (DLRMs) stand out due to their unique characteristics: unlike their compute-heavy counterparts, DLRMs are predominantly bandwidth-intensive due to their large embedding table accumulations~\cite{desai2022trade,shi2020compositional,ke2020recnmp,jain2023optimizing}. This distinction shifts the performance bottleneck from compute capacity to data bandwidth and transfer efficiency. The research community and industry have proposed several hardware-based solutions, such as Process-near-Memory (PNM)~\cite{ke2020recnmp,yang2023pimpr,wilkening2021recssd,sun2022rm,kwon2019tensordimm} and ASIC designs~\cite{hwang2020centaur} to address these challenges. However, these solutions introduce new challenges. Firstly, the ever-expanding size of industrial DLRM models, now surpassing even the most significant Large Language Models (LLMs)~\cite{DLRM19,guo2023software, QuoRemTrick19}, poses a significant challenge to the scalability of PNM and ASIC solutions, due to the limited physical interface on boards. 
Secondly, the PNM-based solutions, by their nature, diverge from standard DRAM protocols~\cite{keeth2001dram}. Consequently, adapting to these memory technologies may demand extensive hardware and software stack modifications~\cite{wilkening2021recssd,shen2024archer,singh2018review,khan2020survey}, elevating development costs and extending the product development cycle. Thirdly, they also introduce resource inefficiency -- with shared memory capabilities at the board level, the PNM solutions can serve limited sockets per board. This limitation may cause redundant data copies within a rack or even across  racks~\cite{kalamkar2020optimizing,wang2022merlin,pumma2021semantic} to facilitate multi-host access, leading to inefficient memory usage and increased latency, despite the advancements like RDMA~\cite{guo2023software,yuan2023rambda}.

Compute Express Link (CXL) technology~\cite{ComputeExpressLink} is rapidly gaining traction in the contemporary datacenter landscape, setting a new standard in the industry. It ensures cache coherence over the PCIe physical layer and introduces memory pooling by using fabric switch~\cite{li2023pond}. This advancement offers enhanced memory scalability and utilization, leading towards a new data processing and management era. Furthermore, recent studies emphasize CXL's capability to operate as a separate and independent memory bandwidth source~\cite{sun2023demystifying,maruf2023tpp}, significantly enhancing the system's overall bandwidth availability. Together, these features provide a robust foundation for accelerating DLRMs at the datacenter scale.

Motivated by these observations, this study leverages the capabilities of the CXL standard (bandwidth/memory expander) and its interconnects to accelerate DLRMs. We present {\bf PIFS-Rec} ({\bf Process-In-Fabric-Switch for Recommendation Systems}), a scalable, near-data processing capability tailored for fabric switch hardware. Focusing on large-scale industrial DLRM inference systems, PIFS-Rec utilizes the scalability of downstream ports~\cite{xconn, ComputeExpressLink} and proximity to memory within the CXL fabric switch~\cite{ComputeExpressLink} to accelerate the embedding table operations. Through minimal hardware and software optimizations within the fabric switch, we extend its capabilities beyond current implementations, including DRAM-based Type 3 memory expanders~\cite{park2022scaling, MicroCXL, SMCXL}. Our design (\S \ref{section3}) enhances existing CXL memory systems by leveraging the ``scalable bandwidth'' of fabric switches to address bandwidth bottlenecks in embedding table accesses of DLRM. This boosts performance through enhanced device-level I/O utilization and parallelization. Additionally, we explore integrating a process-on-a-fabric-switch framework to reduce data movement costs. 

Our main {\bf contributions} in this work are as follows:

$\bullet$ We present results from a characterization study that analyzes recommendation models using real-industrial access traces, production-scale DLRM models, and CXL-ready system. We quantitatively assess the bottlenecks of CXL-enabled memory pooling as well as the potential opportunities it brings. 

$\bullet$ We introduce \textbf{PIFS-Rec}, a scalable near-data processing approach customized for fabric switch hardware. Our optimizations include hardware and software enhancements such as data repacking, snooping mechanisms, on-switch buffer implementation, and optimized compute logic. Additionally, we explore software-assisted page management strategies to enhance the efficiency of the DLRM processing pipeline.

$\bullet$ We use open-sourced industrial DLRM traces to quantify the effectiveness of our optimizations. We find that PIFS-Rec outperforms an existing CXL-based system, Pond~\cite{li2023pond} by 3.89$\times$ and a state-of-the-art comparable design, BEACON~\cite{beacon} by 2.03$\times$ in terms of latency.



\section{Background and Related Works}
\begin{figure}
    \centering
    \includegraphics[width=1\linewidth]{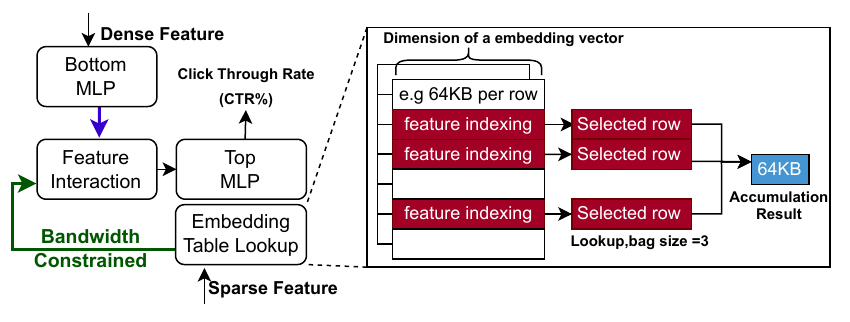}
    \caption{\textbf{End-to-end DLRM pipeline for inference.}}
    \label{fig:dlrmpipeline}
\end{figure}

\subsection{Deep Learning Recommendation Model (DLRM)}
The end-to-end inference in DLRM involves several stages. Initially, the recommendation model is loaded into DRAM. Incoming input queries are grouped into batches, and the necessary dense and sparse features are organized for input into the DLRM. The inference step then processes these batches to generate predictions. In a DLRM architecture, four key stages can be identified: the bottom fully-connected layer (Bottom MLP), embedding lookup, feature interaction, and top fully-connected layer (Top MLP), as depicted in Figure~\ref{fig:dlrmpipeline}. This architecture handles two types of input features: ``dense features" which are continuous personal variables (e.g., age, gender), and ``sparse features" which are categorical (e.g., product IDs, music genres). During the inference process, dense features undergo processing by the Bottom MLP. On the other hand, sparse features are transformed into ``dense latent representation" in the embedding lookup stage. Each feature's value or indices are used to retrieve the corresponding ``embedding vector" from a large table. Embedding vectors can be of different dimensions (e.g., 16B, 32B, etc.) in different setups; high embedding dimensions and the number of embeddings lead to large memory footprints. The outputs from the Bottom MLP and the embedding lookup are combined in the feature interaction layer to calculate interactions before being passed to the Top MLP for determining the click-through rate (CTR).

\subsection{CXL Overview}
\subsubsection{Conventional CXL}
CXL operates as a transaction layer designed for rack-level memory pooling~\cite{li2023pond}, building upon the physical layer of PCIe. PCIe 5.0  supports a data transfer rate of 32 GT/s per lane, translating to approximately 64GB/s when utilizing $16\times$ lanes. As CXL uses PCIe's physical layer, unlike RDMA, it does not require a device's DMA engine or Network Interface Card (NIC). CXL encompasses  three protocols: ``CXL.io'', ``CXL.cache'', and ``CXL.mem''. The CXL.io protocol configures and establishes connections between CPUs and CXL devices. In contrast, the CXL.cache (resp. CXL.mem) protocol enables a device to access the host CPU cache (resp. memory) and vice versa. These protocols enable three types of CXL devices: {\em Type 1} (only cache, e.g., NIC), {\em Type 2} (both cache and memory, e.g., GPU, accelerator, etc.), and {\em Type 3} (only memory, e.g., memory expander).

\begin{figure}[h]
    \centering 
    \includegraphics[width=\linewidth]{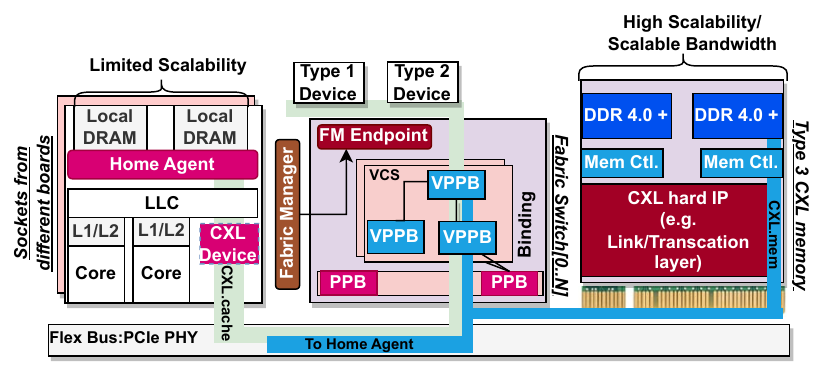}
    \caption{\textbf{Architecture of a CXL-based system. The devices use Flexbus to communicate with the host. The fabric manager configures the Virtual PCI-to-PCI Bridge (VPPB) to control the FM endpoints in the fabric switches. These switches connect all devices within the system. Data leaves fabric switch through PCI-to-PCI Bridge (PPB). 
    }}
    \label{fig:detail_arch_cxl}
\end{figure}
\textbf{Memory Expander.} 
Type 3 devices also known as CXL memory expanders~\cite{SMCXL,MicroCXL}, are designed to increase {\em both} memory capacity~\cite{jang2023cxl} and bandwidth~\cite{sun2023demystifying}. These conventional expanders include DDR memory and rely on the CXL.mem protocol for data storage and retrieval, as well as the  CXL.io protocol for establishing the connections. This protocol enables CPU-to-device memory access, coordinated by a Home Agent (HA) and a CXL controller within the host and device. The HA manages the CXL.mem protocol, presenting memory to the CPU as if it were on a remote NUMA node, allowing direct access via standard load/store commands. 

\textbf{Memory Pooling.} 
CXL enables memory pooling where each host and end device can access any shared, unified memory space connected through CXL. This addresses memory stranding~\cite{li2023pond} and redundancy issues. 
To manage the data access flow within a memory pool, CXL employs an asymmetric cache protocol and introduces a ``Bias Table'' (4KB per table). This table operates in two modes: ``host-bias'' and ``device-bias''. In the host-bias mode, devices accessing addresses within CXL memory need control instructions to ensure data coherence, adding extra overhead. Conversely, in the device-bias mode, the region is locked for the device's exclusive use, preventing access by other hosts. 

\subsubsection{Fabric Switch}
As shown in Figure~\ref{fig:detail_arch_cxl}, In the advancement from version 1.0 to 3.1, CXL introduces the ``fabric switch" in CXL 2.0~\cite{ComputeExpressLink}. Note that the fabric switch is a compulsory component and non-bypass hardware in a multi-node CXL interconnect. Unlike the one-to-one communication of earlier CXL versions (1.0/1.1), CXL 2.0 and later versions facilitate multiple-device communication and interconnectivity. The fabric switch functions as both a memory request dispatcher and a connected device manager. Each device is assigned a {\em cacheID} when recognized by the FM endpoint in the fabric switch.

\textbf{Process-In-Fabric-Switch.} To the best of our knowledge, in the context of CXL, BEACON~\cite{beacon} is the first work that adds compute capability to a fabric switch. Specifically, BEACON integrates ``compute units'' within the fabric switch to harness the processing capabilities close to the data source and the high bandwidth of the downstream port for accelerating genome workloads. Note that BEACON has been developed to accelerate ``genome analysis''. Our analysis reveals several areas where BEACON may not fully maximize the potential benefits of in-switch computation. From a hardware perspective, BEACON's design relies on custom DIMM instructions for CXL memory management, diverging from the established CXL standards. It also needs an additional memory translation logic in the fabric switch which can introduce performance overheads. Moreover, the system is not scalable as it does not support fabric switch scaling. It also does not take advantage of data locality. On the software front,  existing work~\cite{maruf2023tpp,li2023pond,sun2023demystifying,jang2023cxl} suggests that directly accessing CXL memory without careful management can detrimentally affect performance across various workloads, primarily due to higher data access latency compared to local DRAM. BEACON's standalone use of CXL memory, without integrating address interleaving with local DRAM, might result in suboptimal performance. Additionally, it focuses solely on single-host configurations, neglecting the complexities and opportunities brought by multi-host environments.

To address BEACON's shortcomings and maximize the potential of in-switch compute capability in the context of DLRM workloads, we undertake a comprehensive redesign encompassing {\em both}  hardware and software support, establishing a new workflow and a new instruction flow. This effort results in the development of PIFS-Rec, which extends beyond DLRM workloads to cater to a broad range of applications with a focus on enhancing scalability, efficiency, and performance. 


\subsection{Related Works} 
Numerous works address the acceleration (\cite{guo2023software, daghaghi2021accelerating,gupta2020architectural, firoozshahian2023mtia}) and optimization (\cite{gupta2020deeprecsys,yin2021tt, sethi2022recshard}) of DLRMs. 
Software-based approaches focus on techniques like feature-based resource allocation (CAFE~\cite{zhang2024cafe}) and CPU optimizations for pre-fetching and overlapping computation with memory access (~\cite{jain2023optimizing}). Hercules~\cite{ke2022hercules} provides an adaptive scheduler to deploy various DLRM models across the datacenters with heterogeneous devices, considering multiple factors such as power budget, latency requirement, and throughput. In comparison,  DisaggRec~\cite{ke2022disaggrec} explores the deployment of DLRM using hardware resource disaggregation to improve cost efficiency. Note that both of these solutions explore effective scheduling strategies targeting existing servers, while PIFS-Rec is a new hardware acceleration-based solution that targets scalability.

Hardware-based solutions leverage technologies like PIM  \cite{ke2020recnmp,wilkening2021recssd,wang2023ems,wang2023ems,kwon2019tensordimm} for faster data processing within memory and specialized ASICs designed for DLRMs~\cite{hwang2020centaur}.  Furthermore, recent research also explores CXL for both characterization and optimization purposes. For example, Pond~\cite{li2023pond} focuses on memory pooling with CXL to increase scalability;  TPP~\cite{maruf2023tpp} improves CXL system performance with tiered memory page management; and studies like~\cite{sun2023demystifying, kona} explore the potential performance gains from CXL. Our research is centered on leveraging fabric switches in the context of memory disaggregation over CXL.


\begin{figure}[t]
    \centering
    \includegraphics[width=\linewidth]{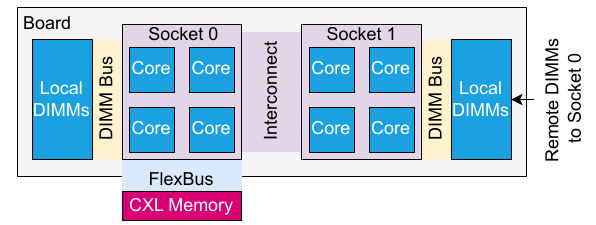}
    \caption{\textbf{A simplified illustration of the production-ready CXL-enabled experiment platform.}}
    \label{fig:experiment}
\end{figure}

\begin{figure}[t]
    \centering
    \includegraphics[width=\linewidth]{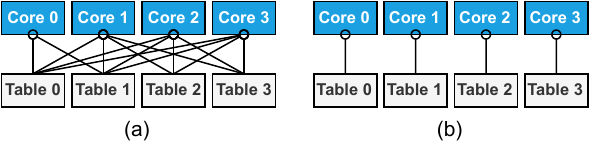} 
    \caption{\textbf{(a) Batch Threading -- each batch is assigned to a CPU core to be processed. (b) Table Threading -- each embedding table is accessed by a CPU core to be processed.}} 
    \label{fig:b_t_threading} 
\end{figure}

\begin{figure*}[t]
    \centering
    \includegraphics[width=\linewidth]{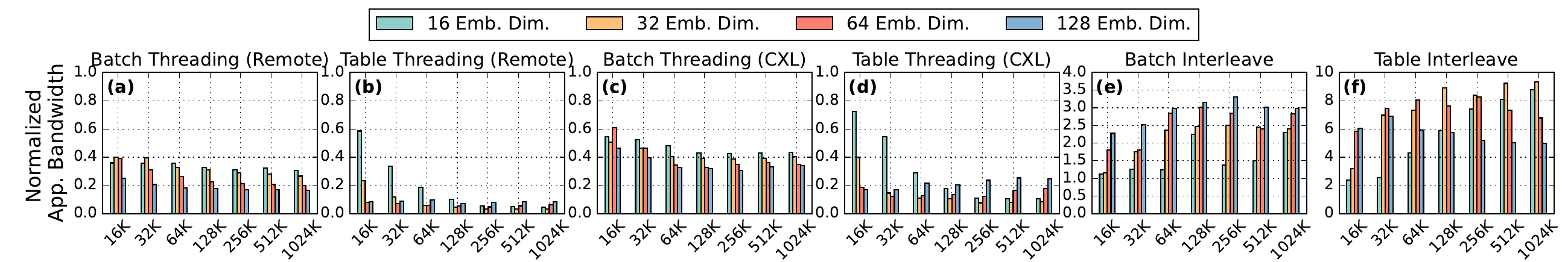}
    \caption{\textbf{The X-axis indicates the embedding table size and Y-axis indicates the normalized application bandwidth. (a)-(b) The addition of CPU sockets can address the scale-up issue of memory-bound embedding table lookup operations at the cost of high-performance overhead. (c)-(d) CXL memory can provide better performance over remote CPU sockets. However, simply replacing CPU-attached memory with CXL memory causes performance overheads during high memory traffic over CXL. (e)-(f) Software interleaving during page allocation improves performance through CXL’s bandwidth expansion.}} 
    \label{fig:charct} 
\end{figure*}

\begin{figure}[t]
    \centering
    \includegraphics[width=\linewidth]{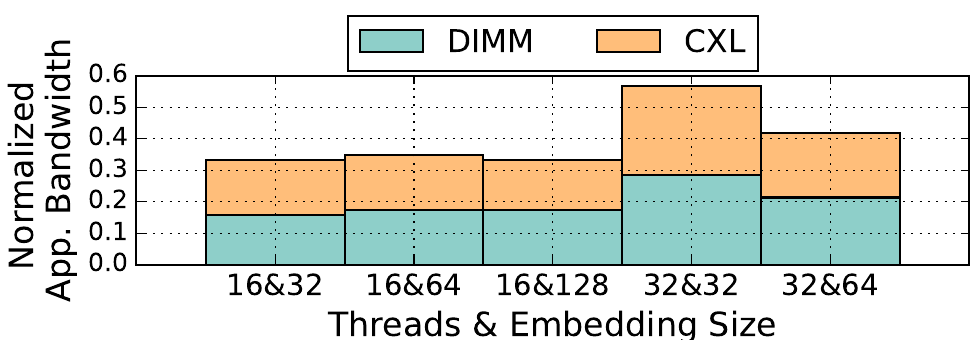}
    \caption{\textbf{The CXL bandwidth contribution to the system with different workload configurations.}}
    \label{fig:morecxl}
\end{figure}

\section{Characterization Study and Motivation} 
The DLRM inference process is both memory capacity and bandwidth bound -- the working set size increases with different parameters (e.g., the number of embeddings, embedding dimension, batch size, and number of tables) while the parallel computation demands high memory bandwidth. To get more memory bandwidth along with larger capacity, one possible solution in today's server system is to add more CPU sockets and populate memory channels in the power of two. However, this restricts flexibility and results in stranded memory resources~\cite{li2023pond}.
As DRAM is a significant driver of infrastructure cost and power consumption~\cite{maruf2023tpp}, its excessive underutilization also leads to high total cost of ownership (TCO). Further, the interconnect between the CPU sockets can be a bottleneck and significantly impact the performance.

To illustrate this, we run the embedding table lookup phase of the DLRM inference process on a dual-socket AMD Genoa system (each socket having 96 physical CPU cores and 12 channels of DDR5 memory that populates a capacity of 768GB), with a representative DLRM trace. As shown in Figure~\ref{fig:experiment}, 
besides adding more CPU sockets, we can also increase the overall system's memory and bandwidth capacity by populating memory channels through CXL interconnects.
Here, CXL memory is enabled through four channels of DDR4 memory resulting in memory capacity of $256$GB. Including the CXL memory, the server has a total memory capacity of $1.8$TB.
Here, we configure the DLRM trace to use 192 tables with a batch size of 1024, and run embedding table look-up operation with different parallelization methods, namely, ``batch threading'' and ``table threading'' (detailed in Figure~\ref{fig:b_t_threading}).

\label{sec:section}
CXL allows flexibility in server memory capacity population, which is restricted during remote CPU socket-based memory capacity expansion. To consume the full memory bandwidth, we must populate and access all the memory channels in the remote socket. While accessing remote socket memory partially, the effective memory bandwidth might be low. For example, Figure~\ref{fig:charct} (a)-(b) illustrates that accessing only 20\% of the whole working set size from a remote CPU socket with DDR5 DIMMs invariably reduces the application bandwidth consumption. This significantly impacts performance, particularly with large embedding dimensions and large numbers of embeddings, where we observe up to 95\% degraded performance. 
In contrast, instead of remote CPU memory, while accessing the same amount of memory (i.e., 20\% of the working set size) over CXL interconnects with DDR4 DIMMs, we can have an enhanced performance of 5--30\% (Figure~\ref{fig:charct} (c)-(d)). Note that CXL-attached DDR4 memory has a low refresh rate over CPU-attached DDR5 memory. Also, CXL memory is CPU-less -- we do not need additional CPUs to expand memory capacity. Consequently, CXL memory can consume less power than remote CPU sockets. Moreover, re-purposing earlier generations of DDR DIMMs can save datacenter TCO while augmenting performance.  

As CXL adds bandwidth to the overall system, it can act as a ``bandwidth expander''  when CPU-attached memory channels are saturated. For DLRMs with large thread counts, batch sizes, embedding dimensions and sizes, both the working set size and memory bandwidth increase significantly.
At some point, the CPU-attached memory channels get saturated and become the performance bottleneck. In such cases, extra bandwidth from CXL can enhance the application throughput. For example,  as shown in Figure~\ref{fig:morecxl}, when we increase the thread count from 16 to 32 and embedding table dimensions from 64 to 128, system bandwidth increases by 43\%. Here, DDR4-based CXL memory improves application throughput by 28.5--38.9\%, compared to the standalone CPU-attached DDR5 memory system. 

The potential of CXL to improve system performance and efficiency is evident. Our findings also highlight several ``limitations'' with the current CXL interconnects. These include the risk of flex bus congestion under heavy memory traffic, increased data access latency (by over 4$\times$ compared to local DRAM~\cite{li2023pond,maruf2023tpp}), and limited bandwidth expansion capabilities. Such constraints can lead to substantial performance degradation in specific configurations, with the observed impact reaching up to 90\% when the bandwidth is saturated. Our analysis of the software front suggests a promising strategy involving the {\em distribution} of embedding tables across available memory tiers. We tried with different interleave ratios and empirically found that, when we allocate 20\% of the total working set size to CXL memory and the remaining 80\% to local DRAM, i.e., we allocate over a 4:1 interleave policy, we get a significant performance improvement (as shown in Figure~\ref{fig:charct} (e)-(f)). We can achieve up to a $9\times$ performance increase over configurations where all memory is allocated to the CXL. Table threading scenarios can offer up to a $1.73\times$ performance boost compared to operations running solely on local DRAM. These results reveal that any method that relies solely on CXL memory (e.g., \cite{beacon}) may restrict performance. Our experiments underscore the significant potential of CXL technology in enhancing scalability and performance. 

Key takeaways from our characterization experiments are:

$\bullet$ \underline{Key Takeaway 1:} While CXL memory enhances system scalability by offering more flexible memory configurations, its data retrieval latency is higher than DRAM. This adversely affects performance. To mitigate this, computation should happen close to memory to minimize the data transfer latency.
 
$\bullet$ \underline{Key Takeaway 2:} The CXL memory can outperform remote CPU socket configurations but requires memory management strategies. Specifically, spreading memory between DRAM and CXL, coupled with careful page management, can significantly reduce performance degradation.

We believe, from a hardware perspective, PIFS is a suitable solution to address these issues. First, placing a fabric switch closer to the memory reduces data movement significantly compared to traditional host-centric models. Second, unlike the PNM/PIM solutions, PIFS does {\em not} require modifications to existing CXL devices, thus maintaining compatibility. We can also optimize {\em both} memory usage and system performance by integrating PIFS hardware with effective page management with DRAMs. Based on these considerations and observed characteristics, we propose {\bf PIFS-Rec}. 



\newcommand{\bluecircleone}{
\tikz[baseline=(char.base)]{
\node[shape=circle,draw,inner sep=2pt,fill=black,text=white] (char) {1};
}}
\newcommand{\bluecircletwo}{
\tikz[baseline=(char.base)]{
\node[shape=circle,draw,inner sep=2pt,fill=black,text=white] (char) {2};
}}
\newcommand{\bluecirclethree}{
\tikz[baseline=(char.base)]{
\node[shape=circle,draw,inner sep=2pt,fill=black,text=white] (char) {3};
}}
\newcommand{\bluecirclefour}{
\tikz[baseline=(char.base)]{
\node[shape=circle,draw,inner sep=2pt,fill=black,text=white] (char) {4};
}}

\begin{figure*}
    \centering
    \includegraphics[width=\linewidth]{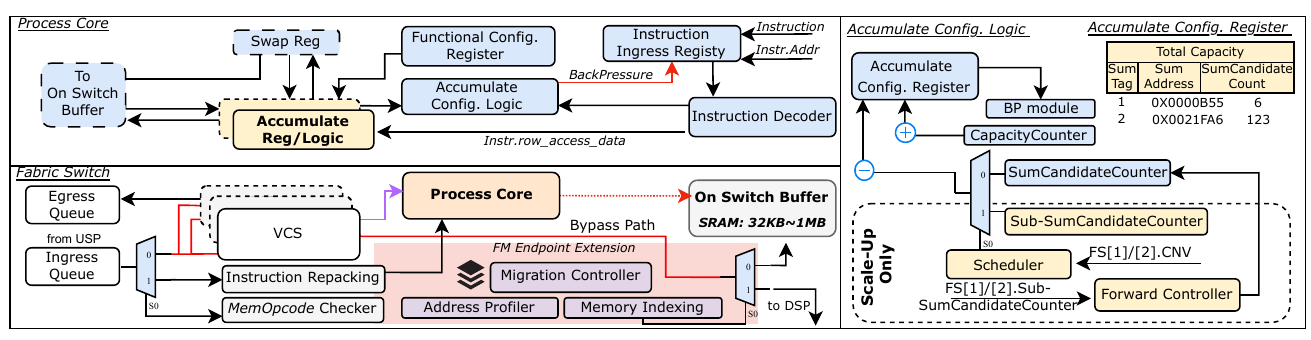}
    \caption{\textbf{Overview of Process-In-Fabric-Switch (PIFS) for DLRMs architecture, which includes Process Core (PC), accumulate configuration logic, accumulate configuration register, and a mechanism for instruction ingress registry.}}
    \label{fig:pifs} 
\end{figure*}

\section{System Design}
\label{section3}
PIFS-Rec scales the embedding tables and accelerates the embedding operations, known as SparseLengthSum (SLS), by leveraging device I/O level parallelism, and facilitates processing near memory by executing computations within a fabric switch. PIFS-Rec features a minimalist hardware architecture with specialized computation tailored to support the SLS family of inference operators. This special-purpose computation logic is localized to the Process Core (PC). Additionally, PIFS-Rec includes an enhanced memory controller (FM endpoint Extension) that extends the functionality of the Fabric Manager (FM) endpoint within the fabric switch. Throughout this paper, we refer to this component as the ``memory controller''.

Figure~\ref{fig:pifs} shows the process flow (\S \ref{sec:processflow}) and a new instruction (\S \ref{sec:instructionmodification}) to minimize modifications to standard CXL-based DRAM devices (Type 3) and mitigate additional overhead. Considering the ``localized nature'' of embedding table access patterns, our framework explores employing an on-switch buffer (\S \ref{sec:buffer}) to enhance overall system efficiency. We also implement an ``out-of-order engine'' (\S \ref{sec:ooo}) to prevent pipeline stalling during DLRM row access accumulation. Additionally, we enhance the software architecture (\S \ref{sec:soft_arch}) through an optimized page management and migration process to complement our hardware design. Furthermore, we discuss scaling up multiple PIFS-Rec interconnections (\S \ref{sec:scaleing}) by introducing multi-layer forwarding and necessary modifications to support this growth. These optimizations draw upon empirical insights from our workload characterization in a CXL hardware-ready memory system and prior research~\cite{DLRM19,ke2020recnmp,jain2023optimizing,wilkening2021recssd}, ensuring a grounded and practical approach to tackle the complexities of modern recommendation systems.

\subsection{Hardware Architecture} 
One of our objectives is to keep the fabric switch lightweight with minimal hardware and software modifications, ensuring cost-effective deployment and compatibility. We make several changes to the conventional fabric switch -- from a hardware perspective, we design the processing core to passively receive instructions from the host and operate exclusively on physical addresses, eliminating the need for a softcore or host presence within the fabric switch. With this, we do {\em not} need to modify the hardware or software on CXL end devices, which facilitates seamless integration to existing Type 3 devices.

\subsubsection{System Overview} 
PIFS-Rec is integrated within the fabric switch, as depicted in  Figure~\ref{fig:pifs}. The PC (Process Core), a hardware component within the fabric switch, facilitates this integration. The memory controller for PIFS-Rec is an FM endpoint extension with an enhanced memory indexing unit. Communication between the host-side CXL controller HA (Home Agent) and PIFS-Rec occurs via CXL-based instructions through the CXL interface using PCIE PHY. PIFS-Rec returns the accumulated embedding table row access results to the host. Regular CXL-based instructions are decoded by the FM endpoint extension and forwarded to the corresponding CXL devices with  the modified instructions. By locating the logic within the fabric switch, PIFS-Rec can issue ``concurrent requests'' to {\em parallel CXL devices} and efficiently utilize bandwidth across {\em multiple memory channels}. The embedding table region is designated as a device-bias region.

\subsubsection{Process Flow}
\label{sec:processflow}
Previous work~\cite{beacon} introduces customized DIMM-based instructions and an independent CXL workflow, diverging from the current CXL protocol standard~\cite{ComputeExpressLink}. To avoid complete hardware and software stack changes, {\em we design the fabric switch from scratch, maintaining compatibility with CXL memory and avoiding major modifications to the CXL host-device control protocol}. 

\begin{figure}
    \centering
    \includegraphics[width=\linewidth]{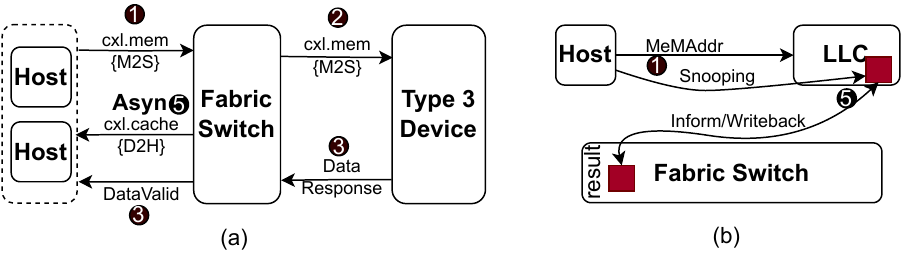}
    \caption{\textbf{(a) Instruction flow of PIFS-Rec.  The valid signal indicates a successful retrieval of data. (b) The host gets the result from the fabric switch.}}
    \label{fig:instrucitonflow}
\end{figure}

\textbf{Instruction Flow.} 
In BEACON~\cite{beacon}, the computational logic within the fabric switch initiates memory requests. However, bypassing the host in a cloud-based inference system presents challenges, as each query might access different row candidates. Hence, the host must relay essential memory address information to the fabric switch for accurate memory access. The PC decodes instructions from the host and issues memory fetch request to specific CXL memory devices. In Figure~\ref{fig:instrucitonflow}, during row accumulation access, the host issues a standard CXL.mem \{M2S\} request~\cite{ComputeExpressLink} to the fabric switch\bluecircleone, while reserving a memory address in the LLC or specific CXL cache region and transmitting it to the fabric switch's process core register. Upon receiving a memory request from the host, the memopcode checker examines the instruction's memory operation (\textit{memOpcode}) field. If the instruction is standard, it bypasses the processing core and is sent directly to the VCS. Otherwise, it is routed to the process core for further handling. After receiving CXL instructions via the interface, the process core decodes the instruction and proceeds with \textit{instruction repacking}. 
This repacking modifies two instruction fields: 
Firstly, for the requests initiated by the host as read requests, \textit{memopcode} is modified to transform them into standard read requests with data directed toward the CXL memory\bluecircletwo. From the host (CPU) point of view, it issues the memory read request, but the actual memory issue request source point comes from the fabric switch. However, the host still acts as a ``monitor'', that is, if the address is polluted or invalid, the host will realize it and inform the application or runtime. Secondly, the repacking alters the \textit{SPID} (the ID of device that initiated the request) from the host to the fabric switch, ensuring that the retrieved data are stored in the fabric switch.  Once the data are retrieved from the memory and sent to the  fabric switch\bluecirclethree, the process core  dispatches a control signal back to the host\bluecirclethree, indicating successful data retrieval.

\textbf{Asynchronous Communication.~} 

As mentioned earlier, the fabric switch's processing core initiates data accumulation. When the embedding tables interleave between local DRAMs, remote DRAMs, and Type 3 devices, the host computes the \textit{SumCandidateCounter} for each request to accumulate rows. The host first identifies all related row vector candidates' memory addresses (using the data\_ptr() API in PyTorch). It then uses the memory addresses to determine the location of each row vector, checking whether it is in the local DRAM or elsewhere (using the move\_pages() API in numactl). Subsequently, it calculates the \textit{SumCandidateCounter} by tallying the number of vectors not stored in the local DRAM. Note that \textit{SumCandidateCounter} is configured into a fabric switch using the instruction. Specifically, the PC decrements the counter by 1 each time it accumulates a row candidate. The process is considered complete when the \textit{SumCandidateCounter} reaches 0. Upon the completion of the accumulation process, the accumulated result is transmitted to the previously reserved memory address of the host with CXL.cache \{D2H\}\bluecirclefour  through the egress queue. The host continuously monitors (snoops) the designated address using the standard CXL snooping mechanism. Upon detecting a change in the memory location, it recognizes that the data at this location represents an accumulated result. It then retrieves the accumulated data for further processing.  
\begin{figure*}[t]
    \centering
    \includegraphics[width=\linewidth]{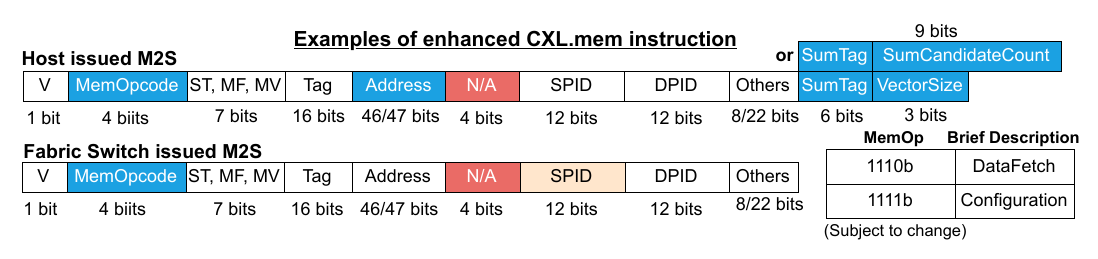}
    \caption{\textbf{Blue chunks indicating modified/added fields and SPID modification by the fabric switch.}}
    \label{fig:instruction}
\end{figure*}

\newcommand{\inversecirclethree}{
\tikz[baseline=(char.base)]{
\node[shape=circle,draw,inner sep=2pt,fill=black,text=white] (char) {3};
}} 
\subsubsection{Instruction Modification}
\label{sec:instructionmodification} 
To implement the described mechanism (\S~\ref{sec:processflow}), modifications have been made to the instruction set (CXL 3.0), as depicted in Figure~\ref{fig:instruction}. Specifically, the memory opcode within the Request (Req) instruction serves dual purposes: it can either initiate a request for row vector data or configure the Accumulation Configuration Register (ACR). For row vector data access, the instruction includes a \textit{sumtag} that designates the accumulation cluster to which it belongs and specifies the vector size. Conversely, if the instruction is intended for configuration, it conveys the number of row vectors needed for a particular row accumulation (\textit{SumCandidateCount}). In this case, the  address field is re-purposed to specify the location reserved for the accumulated result, which is then set within the ACR. The  minimum data granularity managed is 16B, while the row vector size can vary~\cite{DLRM19} from 16B to 64B or 128B ~\cite{jain2023optimizing,ke2020recnmp}. The \textit{vectorsize} field indicates the number of data chunks to form a row access, supporting 8 different row vector size configurations with 3-bit width using binary coding. 
Considering the CXL standard's slot size limitation of 16 bytes~\cite{ComputeExpressLink}, weight (since FP32 for weight has 32 bits) and other extra information are allocated within the data slot field. When a memory fetch based instruction arrives at the PC, it is stored in Instruction Ingress Registry (IIR). New data arriving from the CXL memory to the fabric switch is indexed in the IIR, and the corresponding instruction is retrieved by comparing the \textit{address} field. This instruction is then forwarded to the instruction decoder, which sets up the ACR based on instruction's fields. Each new row accumulate request from host is assigned a \textit{sumtag}; each new request increases the capacity counter; and each finished request decreases it. If the ACR hits its capacity limit \textit{CapacityCounter}, the system imposes back-pressure on the upstream modules until space is freed. This cycle continues until all data elements are processed, culminating in the dispatch of the result to the host.

\subsubsection{On-Switch Buffer}
\label{sec:buffer}
The on-switch buffer in PIFS-Rec utilizes on-chip SRAM and acts as a ``cache'' to store frequently accessed ``hot content''. Unlike prior works that use buffers for queue management or traffic shaping~\cite{addanki2022mars,addanki2022powertcp, apostolaki2021fb,yu2021buffer}, our buffer is specifically designed to exploit the temporal locality observed in specific embedding tables where vectors are frequently reused~\cite{ke2020recnmp}. Conventional  prefetching~\cite{jain2023optimizing} strategies are less effective due to the irregular, time-wise relationship patterns exhibited by row accesses, potentially degrading system performance by consuming available bandwidth budget or displacing vectors prematurely. RecNMP~\cite{ke2020recnmp}, a PNM-DIMM-based solution, explored DIMM caching to reduce latency by leveraging data locality. We integrated an on-switch  ``buffer'' to exploit the reuse of embedding vectors. Fetching a single address from memory pools can take up to 270 ns, with approximately 37\% attributed to frequent CXL I/O port transfers and retimer delays, as per profiling~\cite{li2023pond}. This reduction in latency is achieved by minimizing wire transfers and reducing CXL I/O port overhead~\cite{li2023pond}. Distinct from traditional strategies like LRU or FIFO, PIFS-Rec employs a strategy, Hottest Recording (HTR), akin to RecNMP~\cite{ke2020recnmp}. An address profiler logs and ranks frequently accessed row vectors, curating the cache to retain highest-priority candidates based on access frequency. Managed by the FM endpoint extension on the fabric switch, this memory region is inaccessible and unmanageable by the host.

\subsubsection{Out-of-Order Accumulation} 
\label{sec:ooo}
In PIFS-Rec, the accumulation operations are processed in the accumulate logic unit. Here, we optimize existing data management solutions on computational logic, leveraging insights from previous research by~\cite{ke2020recnmp,kwon2019tensordimm,wilkening2021recssd}. In scenarios involving multiple hosts or devices, batch requests from various hosts can trigger numerous accumulation requests to different devices. However, access congestion~\cite{Dillow2011IOCA} at frequently-used memory I/O ports may cause delays in the arrival of row data in the embedding table, as observed by~\cite{liao2023rio,app10124341}.

\textbf{Eliminating Hardware Stalls.}
We do not {\em solely} rely on hardware parallelism such as deploying multiple Near Data Processing (NDP) units~\cite{beacon}, due to two main constraints. Firstly, the extensive computational logic required on the fabric switch demands significant amount of resources, limiting scalability. Secondly, the system's throughput is ultimately constrained by the number of parallel compute units, potentially leading to stalls once this limit is reached. To overcome these limitations, we introduce an ``out-of-order'' compute approach, enabling immediate data processing upon arrival of the same accumulation request. In case of incoming data corresponding to a different request, the system transfers the accumulated intermediate result from the accumulation register to a swap register during the first half of the clock cycle, allowing for processing of the new data in the subsequent half. The shared swap region approach among multiple processing cores and accumulation logic ensures efficient data handling. Note that the SRAM in the switch buffer can also contain the intermediate result while the swap register is full. However, accessing data from the SRAM in the switch buffer requires at least two clock cycles, potentially causing stalls. In our current approach, we make this function configurable by configuring the Functional Configuration Register (FCR). 

\raggedbottom
\subsection{Software Architecture}
\label{sec:soft_arch}
From our characterization study (\S \ref{sec:section}), we find that for DLRM, proper utilization of all the available memory channels simultaneously can provide the optimized performance in a CXL-enabled system. 
Considering this, our software architecture incorporates the following design principles --  
\textbf{(a)} as CPU-attached local memory node has the lowest access latency and comparatively high bandwidth over CXL-memory, hottest or most frequently accessed pages should reside on the local memory tier;
\textbf{(b)} When we need to access CXL memory, if we can spread the memory across multiple CXL nodes, then we can parallelize better and utilize the bandwidth across all the channels;
\textbf{(c)} As migration of pages is a widespread event in a tiered memory system, optimizing that software feature can significantly enhance the overall system's performance.

\subsubsection{Page Granular Access}
\label{sec:page}
In DLRM, the dimension of an embedding vector can be very small (e.g., typically, ranges between 16--128B).
As CXL supports cache-line granular access, we can consider the embedding dimension to be the granularity of memory access and efficiently identify the hot-cold rows to perform fine-grained vector embedding management.
However, the metadata management overhead will be high.
On the other hand, a single OS page (e.g., typically, a 4KB-sized page) can contain multiple row vectors (e.g., 256 embeddings of 16B size).
It is possible that all the embeddings within a page may not be accessed simultaneously, which will cause amplification of data movement.
However, even with this caveat, in our system, same as previous work~\cite{li2023pond,maruf2023tpp,sun2023demystifying}, we opt to manage memory placement at page-granular as page-granular metadata management and migration is supported and compatible with the current OS. Hot-cold detection also happens on page-granular.  

\subsubsection{Global Hotness Detection}
\label{sec:em}

\begin{figure}[t]
    \centering
    \begin{minipage}{\columnwidth}
        \includegraphics[width=0.9\linewidth]{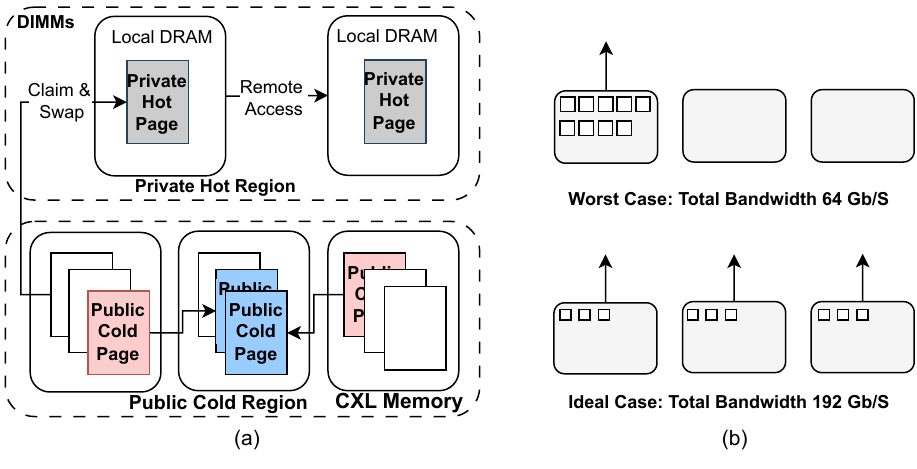}
        \caption{\textbf{(a) Page migration and management. The socket on same board can access remote DIMM using the board-level interconnect, but it needs fabric switch to access remote DIMM on another board. (b) In the worst case, memory requests are not localized across devices. }}
        \label{fig:thread}
    \end{minipage}
\end{figure}

We provide a unified memory architecture where all the hosts can access pages across the system. 
When a host accesses a page frequently, we identify it as the hottest one and put it to its local DRAM (we call it ``Private Hot Region") (Figure~\ref{fig:thread} (a)).
As CXL has higher latency (around 100ns extra over local DRAM), we put the relatively cold pages in the CXL memory address space (we call it ``Public Cold Region"), which is shared between all the connected hosts.
To identify the global page temperature, each host monitors the access frequency of a page across all devices -- the most frequently accessed pages within a device are categorized as ``hottest'' while the least frequently accessed pages are categorized as ``coldest''. 
After generating all the device's page heatmaps, the host compares them.
Therefore, it finds the most frequently accessed pages across all devices and stores them in its private hot region.
If a host identifies a page already designated as a private hot page by another host, it selects its next most frequently accessed page as its private hot page. 
Remote hosts access memory from another host's private hot region over the flex bus, incorporating an accumulation process within the fabric switch. 
If a host retrieves a row vector from local memory, accumulation happens locally, although it is capable of receiving (but only partially processing) the accumulated results. 
Every host periodically reclassifies hot private pages as public cold pages if their access frequency exceeds the least accessed private hot page's access frequency by more than ``cold\_age\_threshold" (by default, 20\%).

\begin{figure}[t]
    \centering
    \includegraphics[width=\linewidth]{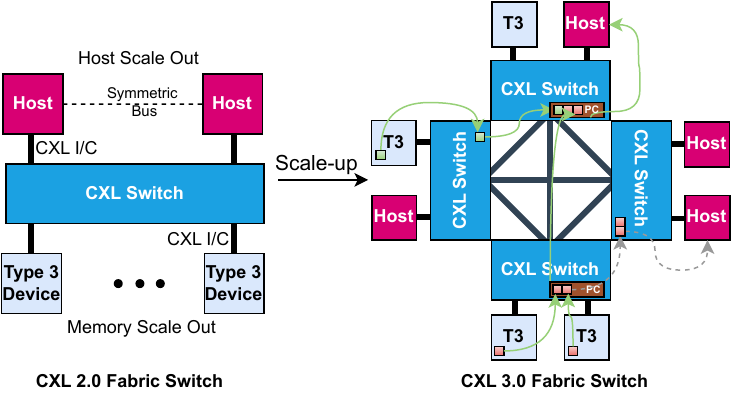}
    \caption{\textbf{From CXL 2.0 to 3.0. PIFS-Rec supports scale-up with multi-host scenarios (T3: Type 3 memory devices).}}
    \label{fig:scale-arch}
\end{figure}

\subsubsection{Embedding Spreading for Bandwidth Optimization}
\label{sec:embeddingspreading}
In our architecture, we address the potential bottleneck scenario due to disproportionate demand on specific memory devices. 
As illustrated in Figure~\ref{fig:thread} (b), despite dedicated processing threads and specialized task allocation, the system's bandwidth may not be fully optimized if a particular memory node consistently handles most data requests. 
To address this, we introduce a simple yet effective adaptive ``page migration strategy'' to ensure the maximal utilization of channel capabilities across the system. The objective is to {\em redistribute} the workload more evenly across the available memory nodes to alleviate the bottleneck and optimize the bandwidth usage. 

When we place the cold pages in the "Public Cold Region," we initially spread them across the CXL memory nodes through the interleave policy. 
At a later point, if a CXL memory node becomes warm, i.e., the memory access count for a node exceeds the average access count for other nodes by ``1 - migrate\_threshold" (by default, 35\%), we initiate the page redistribution process for that particular memory node.
This process entails transferring the most accessed pages from the overburdened memory node to the least accessed one.
If the destination node is out of capacity, we also move the coldest page of that device to the overburdened memory node. 
Therefore, the page with the second-highest access frequency on the overburdened node becomes the new hottest page for that node.
Similarly, if the cold page is moved, the coldest page of the least frequently accessed memory node also gets redefined. 
We re-iterate the procedure across all the memory nodes until the access frequency gets balanced.

\subsubsection{Optimization in Page Migration}
\label{sec:p2p}
Existing work~\cite{ren2021sentinel,choudhary2017critical,maruf2023tpp} usually focuses on page-level migration due to software compatibility and OS support. 
However, page migration during the live-on inference system can stall query processing due to migration overhead and data inaccessibility. 
When a page is migrated, the OS typically marks it as non-accessible (page block). 
In that case, for a row vector dimension of 64B, a 4KB-sized page migration will block access to all the 64-row vectors residing within that page.
To address this, although we use page-granular memory management to reduce metadata overhead, during migration, we leverage CXL's cache-line granular memory access feature.

We enhance the peer-to-peer (P2P) communication mechanism~\cite{ComputeExpressLink}, with the support of the ``Migration Controller'' (MC) in FM endpoint extension. Once the OS triggers migration, instead of copying the whole page, the host migrates in cache-line granularity. So, when the page is migrated, locking on a particular cache line cannot restrict the accessibility of the remaining cache lines. During this process, the cache-line is not stored back in the secondary memory or returned to the host; instead, it is stored in a temporal location in the switch (cache-line block). This  optimization reduces the overhead by up to $5.1\times$ over the OS's page-granular migration process.  

\subsection{Fabric Switch at Scale}
\label{sec:scaleing}
As shown in Figure~\ref{fig:scale-arch}, fabric switches play a crucial role in scaling out, connecting hosts and devices into a ``unified fabric'' for coherent memory sharing and device communication. In a scaled-out CXL environment, multiple hosts connect to a non-tree-shaped CXL fabric switch, facilitating connections to shared Type 3 memory devices or other hosts. This configuration enables a ``distributed computing'' model, distributing computational and memory resources across multiple nodes. Notably, while \cite{beacon} focuses on single fabric switch computation, our work demonstrates how multiple fabric switches can communicate and collaborate.

\subsubsection{Multi-layer Instruction Forwarding}
\label{sec:multilayer}
In a simplified scenario, we assume each fabric switch has a process core. PIFS-Rec enables vector accumulation to be executed directly on the remote fabric switch (close to the Type 3 memory device), thereby conserving a significant network interconnect bandwidth. Consequently, instruction repacking is confined to the remote fabric switch that handles its local memory. The fabric switch tracks the row candidate requests sent to and received from other fabric switches. Its scheduler reads the \textit{sumtage} field and corresponding memory address, recording the number of memory requests for specific row access to each remote fabric switch, and sends the new \textit{Sub-SumCandidateCount} to replace \textit{SumCandidateCount}. This mechanism ensures the sanity of data exchange and accumulation processing. When a fabric switch positioned near the local host (the host issues the row access request) receives results from a remote fabric switch, there is a possibility that some candidates from other nodes have not yet arrived. The forward controller in the fabric switch monitors the \textit{sumcandidatecounter} to address such situations. The remote fabric switch transmits its \textit{sub-sumcandidatecounter} back to the local switch, which then uses this information to determine whether to forward the accumulated result to the host (once all candidates have been processed), wait for missing data from other nodes, or discard the result if errors occurred during data transfer. 

\subsubsection{Versatility}
The framework can also work with fabric switches without processing cores. During the initial setup and configuration phase, the local fabric switch must identify remote fabric switches' lack of processing capabilities. The scheduler will read a 1-bit \textit{CNV} (Compute Node Valid) field for each fabric switch during the configuration process. If it is determined that a remote fabric switch lacks processing power, the local fabric switch will undertake all operations and instruction repackaging tasks by itself. 

\subsection{Programming-Related Aspects} 
PIFS employs an easy-to-use, OpenCL-like heterogeneous computing programming model similar to the one adopted in the previous work~\cite{ke2020recnmp}. More specifically, PIFS-Rec provides SLS APIs that can be called from user-space, allowing integration with mainstream frameworks like PyTorch. Users can utilize this function call to accelerate specific types of DLRMs and all ML models that require sparse network embedding table accumulation. We aim to reduce the programmer's burden by {\em abstracting} as much information as possible. When allocating memory, users must supply the embedding table file as a parameter, along with the number of embeddings and vector size. Additional parameters, including batch size, indices, offset, or length, are also required. After receiving the information, the PIFS kernel allocates memory space from the CXL memory pool using \textit{numactl} mapping information. The allocated memory region will be implicitly defined as host-biased using the OpenCL API, e.g., \textit{clEnqueueSVMUnmap}. Each function call also allocates an address that points to the output result and pins the address. It generates the embedding table iterative accumulation codes using parameters such as length, batch size, and indices. Simultaneously, the host allocates an embedding table based on the CXL-recognized memory space. The CXL-supported CPU compiles and generates CXL instructions with the corresponding \textit{MemOpcode} field by reading the generated PIFS kernel code. The fabric switch then begins to receive the instruction and passively starts the computation. Meanwhile, a daemon process on the host starts snooping the result address and monitoring the process's integrity for each called function. While instructions are being processed in the fabric switch, the memory indexing function directs data to the corresponding devices and retrieves the data. When page migration is triggered and certain pages are mapped to CXL memory, the bias table flip function hooks into \textit{page migration()} and marks it as a device-biased page using APIs (e.g.,  \textit{clEnqueueSVMMap}), informing the host and the fabric switch, changing the candidate count number, and indexing data to a new location (recall that the rest of the workflow is described earlier in Section (\S \ref{sec:processflow})). 

\section{Discussion}
{\bf Workload Generality.} Our proposed framework for PIFS is highly adaptable, facilitating its usage across various practical workloads (e.g. Kv aggregation~\cite{MongoDB2024}, Mapreduce~\cite{dean2008mapreduce} ). This adaptability is achieved by substituting the compute logic, and DLRM-specific registers with components tailored to the new workload. Unlike BEACON\cite{beacon}, which introduces optimizations specific to genome analysis, our proposed optimizations are designed to be applicable across new workloads without modifying the proposed instruction flow and process flow.

{\bf Hardware Compatibility.} PIFS-Rec is designed to {\em complement, not replace}, PNM/PIM-based approaches. PIFS-Rec is compatible with hardware that includes embedded cores on DIMM, such as RecNMP~\cite{ke2020recnmp}, which addresses bandwidth limitations by using intra-DIMM bandwidth. Integrating such technologies increases operating system complexity, demanding runtime and compiler adaptations for CXL-based heterogeneous computing. This integration requires significant software and hardware modifications, including the adoption of CXL-compatible and core-enabled DIMMs. Addressing these challenges will require extensive research in operating system support, memory management, and hardware innovations.


\begin{table}[ht]
\centering
\scriptsize 
\setlength{\tabcolsep}{6pt} 
\footnotesize 
\caption{\textbf{Model parameters in the scope of this study.}}
\label{tab:t1}
\begin{tabular}{@{}lccccr@{}} 
\toprule
\textbf{Name}  & \textbf{Emb. Num} & \textbf{Emb. Dim}   & \textbf{Bottom-MLP} & \textbf{Top-MLP} \\
\midrule
RMC1           & 16384              & 64               & 256-128-128         & 128-64-1         \\
RMC2           & 131072             & 64               & 1024-512-128        & 384-192-1        \\
RMC3           & 1048576            & 64               & 2048-1024-256       & 512-256-1        \\
RMC4           & 1048576            & 128              & 2048-2048-256       & 768-384-1        \\
\bottomrule
\end{tabular}
\end{table}

\begin{table}[h]
\centering
\label{tab:DDR5_configuration_timing}
\caption{\textbf{Hardware configuration.}}
\label{tab:t2}
\scriptsize 
\setlength{\tabcolsep}{35pt} 
\begin{tabular}{@{}ll@{}}
\toprule
\textbf{DRAM Configuration} &  \\ \midrule
DIMM Capacity      & 64 GBs per DIMM \\
DIMM Channels/Ranks     & 4/2 \\
Frequency (MHz)    & 4800 \\
Timings (CL-RCD-RP-RAS) & 28-28-28-52 \\
tRC/tWR/tRTP       & 79/48/12 \\
tCWL/nRFC1/tCK\_ps & 22/30/625 \\\hline
\textbf{CXL Configuration} &                 \\\hline
Fabric Switch Downstream Ports:      & 64GB/s $\times 16$ \\
Fabric Switch Buffer R/W Speed        & 0.91-4.19 ns/0.91-4.17 ns \\
CXL Access Penalty over DRAM: & 100 ns~\cite{maruf2023tpp}\\
\bottomrule
\end{tabular}

\end{table}

\section{Implementation and Evaluation} 
\label{section4}

\begin{figure*}[h]
    \centering
    \includegraphics[width=\linewidth]{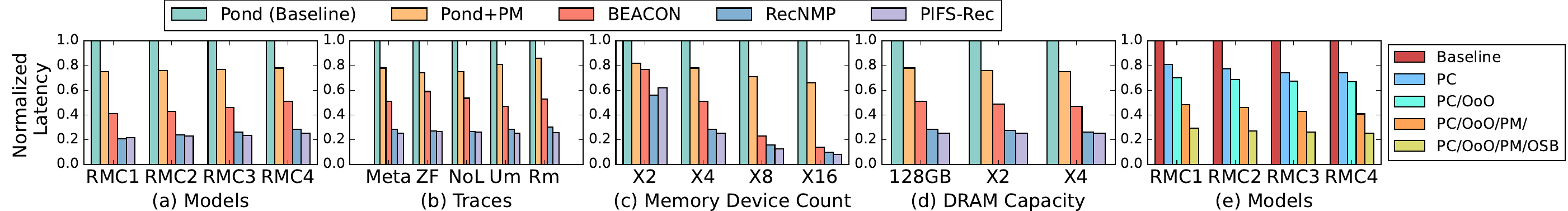}
    \caption{\textbf{The performance of systems with different: (a) models, (b) types of traces (ZF: Zipfian, NoL: Normal, Um: uniform, Rm: Random), (c) memory devices, e.g., X2 means two memory devices, (d) DRAM size, e.g., X2 means 256 GB DRAM, and (e) ablation study: the PC (processing core) is PIFS-Rec specified. The plot uses min-max normalization.}}
    \label{fig:plot14}
\end{figure*}
\subsection{Setup} We adhered to the methodology outlined in a previous study~\cite{ke2020recnmp,beacon}. As described in Table~\ref{tab:t2}, we conducted cycle-level memory simulations using Ramulator 2.0~\cite{luo2023ramulator2} for a detailed evaluation of PIFS-Rec. We wrapped Ramulator 2.0 into our simulator; the top module includes a cycle-accurate processing logic for Process Core, FM Endpoint Extension, Instruction Repacking, and Memopcode Checker and with a top-module clock tick period of one ns/clk. We introduced additional latency (ticks) for data directed to the CXL memory to accurately simulate performance impacts, considering its inherently higher access latency than on-switch DRAM (refers to Table~\ref{tab:t2} ). We use the open-source Meta traces~\cite{metatrace} and models (Table~\ref{tab:t1}) for reproducibility. A ``lookup table'' was developed to facilitate address indexing and mapping logic, directing the memory footprint to either CXL memory or an on-switch buffer. This table is used to record memory access and I/O patterns. Our comprehensive latency evaluation considered several critical factors, such as the additional DRAM cycles required for initializing accumulation counters, the latency introduced by the fabric switch, and the time needed to transfer the final computed sum back to the host system; we extracted the performance from top module synthesis.

\subsection{Baselines}
We selected several previous works as ``baselines'' to highlight the state-of-the-art in various aspects of memory pooling and processing architectures. Pond~\cite{li2023pond} introduces a straightforward CXL-based memory pooling approach with OS support, emphasizing simplicity in design. We add our PM (page management) optimization to Pond, denoted as ``Pond + PM'' to demonstrate the performance of software optimization independently. In comparison, BEACON~\cite{beacon} presents the PIFS architecture to accelerate DNA computation. In our evaluations, we modified the compute logic only to process vector accumulation. Since our main workload is DLRM inference, we implemented the BEACON (BEACON-S) \textit{without} algorithm-specific optimizations. RecNMP~\cite{ke2020recnmp} is a DIMM-based hardware solution that accelerates SLS operations. We implemented the design using their computational hardware configuration with our memory setting. We used a fixed amount of 128GB local DRAM, and memory addresses exceeding this amount will be mapped into CXL regions. Even though BEACON does not support DRAM and CXL interleaving, we still used reduced DRAM latency to access the corresponding 128 GB for BEACON.

\subsection{Evaluation}
To demonstrate the performance benefits of our proposed design for Sparse Length Sum (SLS) operations, we employ various memory devices across different models. Specifically, we utilize four memory devices with default parameters: 8 per batch, RMC4 model, page swap threshold 12\%, embedding migration threshold 35\%, and 512 KB buffer size. We evaluate the performance by using the total ticks used to process the traces and use min-max normalization.

\begin{figure*}
    \centering
    \includegraphics[width=\linewidth]{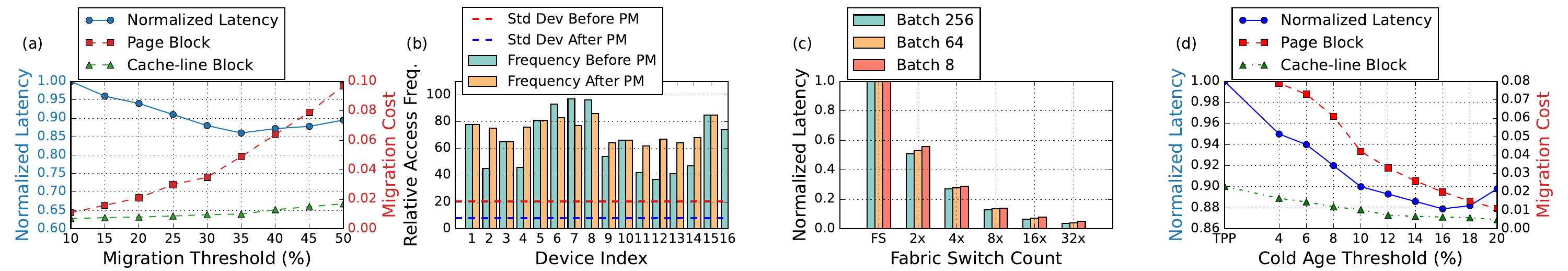}
    \caption{\textbf{Performance Comparison of (a) Embedding migration strategy across different thresholds, displaying normalized latency on SLS-operations (blue). (b) Embedding migration across devices, comparing IO access frequency before and after page migration. (c) Instruction forwarding across various fabric switches and batches. (d) Page swapping strategy for private hot and public cold pages at different thresholds. For (a) and (d), the right Y-axis is a normalized overhead for page block (red) and cache-line block (green). The migration costs are with respect to the total latency.}}
    \label{fig:plot4-8}
\end{figure*}

\subsubsection{HW/SW Co-Evaluation}
Figure~\ref{fig:plot14} (a) presents the performance results with different schemes. Pond, which integrates standard CXL support, demonstrates the lowest performance among the evaluated systems. This is anticipated because, although CXL provides an increase in bandwidth over traditional DIMM-based systems, it is hindered by significant data retrieval latency. As the workload size increases from RMC1 to RMC4, the number of pages mapped to CXL increases. Consequently, the latency increases in all approaches. However, PIFS-Rec has the lowest latency for all these workloads. On average, PIFS-Rec outperforms Pond by 3.8x, Pond + PM by 3.5x, RecNMP by 8.5\%, and BEACON by 1.94x across all the workloads. 
On the other hand, the potential for bandwidth scalability through connecting multiple memory devices via a fabric switch has yet to be fully realized due to the lack of embedding table migration, limiting the system's ability to exploit the increased bandwidth effectively. For the largest model (RMC4), PIFS-Rec achieves only about 11\% improvement over RecNMP because the latter performs data fetch with bank-level parallelism. Also, PIFS-Rec achieves 3.89$\times$, 3.57$\times$ and 2.03$\times$ improvements over Pond, Pond + PM and BEACON, respectively, due to its intelligent page management and optimized device-level parallelism.

\subsubsection{Generality}
In addition to the Meta traces~\cite{metatrace}, we have conducted experiments with synthetic traces to cover a large spectrum of DLRM workload scenarios. Note that Meta traces primarily reflect workload distribution, particularly in Meta's implementation of DLRM, which may not comprehensively represent the diversity of DLRM workloads. Our synthetic traces emulate various distribution types based on the access candidates observed in the Meta traces. As depicted in Figure~\ref{fig:plot14} (b), our findings indicate that the uniform distribution yields the most favorable performance since it creates a perfectly balanced distribution of embedding table accesses across devices. This distribution strategy results in a 1.1$\times$ improvement over RecNMP. Conversely, the Zipfian distribution is identified as the least effective, yielding only a 2\% performance enhancement over RecNMP. Without the help of hardware support to mitigate the bandwidth  bottleneck, Pond + PM improves over the baseline by only 21\%, on average. PIFS-Rec achieves improvements of 2-2.2$\times$ over BEACON and 3.8-3.9$\times$ over Pond, underscoring the importance of careful memory mapping and maximized I/O parallelism. 

\subsubsection{Ablation Study}
We explore several optimizations encompassing both hardware and software enhancements. The results in Figure~\ref{fig:plot14} (d) reveal that adding processing cores yields a modest 26\%  improvement compared to Pond, partially utilizing the high bandwidth. Incorporating out-of-order processing (\S \ref{sec:ooo}) provides at most 7.3\% enhancement due to the elimination of cycle stalling. We observe a performance boost from page management (\S \ref{sec:em},\S \ref{sec:embeddingspreading}), resulting in around 27\% improvement due to optimized memory access and better device-level parallelism. On-switch buffering (\S \ref{sec:buffer}) with PIFS effectively mitigates CXL's high retrieval latency, resulting in an additional 15\% improvement over Pond. Combining out-of-order processing with page migration optimizes I/O parallelism and minimizes stall time. We analyzed the impact of varying on-switch buffer capacities on performance in the following section(\S \ref{sec:capacityreslt}).

\begin{figure}[h]
    \centering
    \includegraphics[width=1\linewidth]{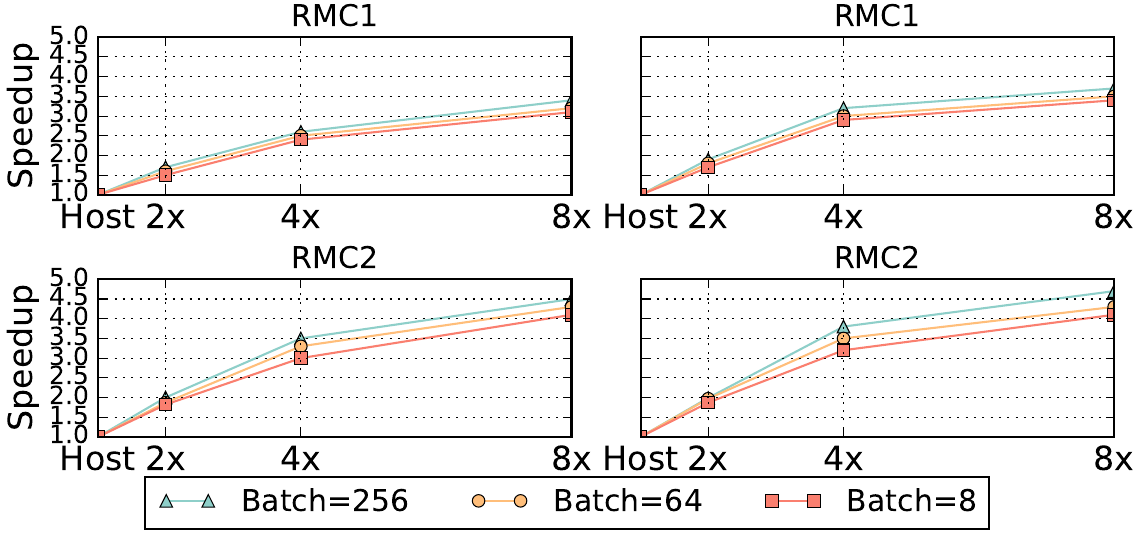}
    \caption{\textbf{Speedup of PIFS-Rec with different numbers of hosts sending concurrent requests under different models.}}
    \label{fig:batch_host}
\end{figure}

\subsubsection{Scalability}
Previous studies~\cite{beacon}, the above discussions, and our experimental analysis collectively confirm that CXL memory pooling enhances scalability compared to DIMM-based solutions. 
We divide the trace file region evenly across memory devices. In Figure~\ref{fig:plot14} (c), the performance can be reduced with limited hardware setup, likely due to constrained optimization opportunities for I/O performance and inherently poorer device-level parallelism (compared to bank-level parallelism). 
Nevertheless, as the hardware inclusion expands, our design demonstrates superior performance in latency, achieving approximately a 12.5$\times$ improvement over Pond, 8.3$\times$ improvement over Pond + PM, a 22\% improvement on RecNMP when there are 16 memory devices. We conduct a sensitivity study by increasing the local DRAM capacity and find that PIFS-Rec still performs the best. Here, the DRAM capacity plays a relatively minor role in shaping the performance. Specifically, 256GB and 512GB DRAM budget result in average performance improvements of 4\% and 6\% compared to 128GB DRAM configuration. This limited effect of DRAM capacity is due to two main reasons -- the model size is in the several terabytes range and the primary bottleneck is memory bandwidth. Therefore, increasing memory capacity alone cannot alleviate the issue of bandwidth saturation.

To estimate the improvements in end-to-end inference latency with multi-host and multi-batch cases, we calculate the speedup by {\em weighting} the speedup of both SLS and non-SLS operators. In Figure~\ref{fig:batch_host}, the performance enhancements due to PIFS-Rec vary with batch size. As the time spent in accelerated SLS operators grows, the model-level speedup increases with larger batch sizes. In RMC4, with the number of hosts increasing from 2 to 8, the performance improves by 1.9--4.7$\times$. With the support of the multi-layer instruction forwarding strategy (\S \ref{sec:multilayer}), We illustrate the latency improvements over different fabric switch counts in Figure~\ref{fig:plot4-8} (c). Assuming each fabric switch has one local CXL memory and one host. These fabric switches are fully connected and we add an extra 100 ns latency when data needs to be transferred between them. The results with RMC4 indicate that as the fabric switch count is increased from 2$\times$ to 32$\times$, the latency improves by 1.8--20.8$\times$ in the largest batch.

\begin{figure}[t]
    \centering
    \includegraphics[width=1\linewidth]{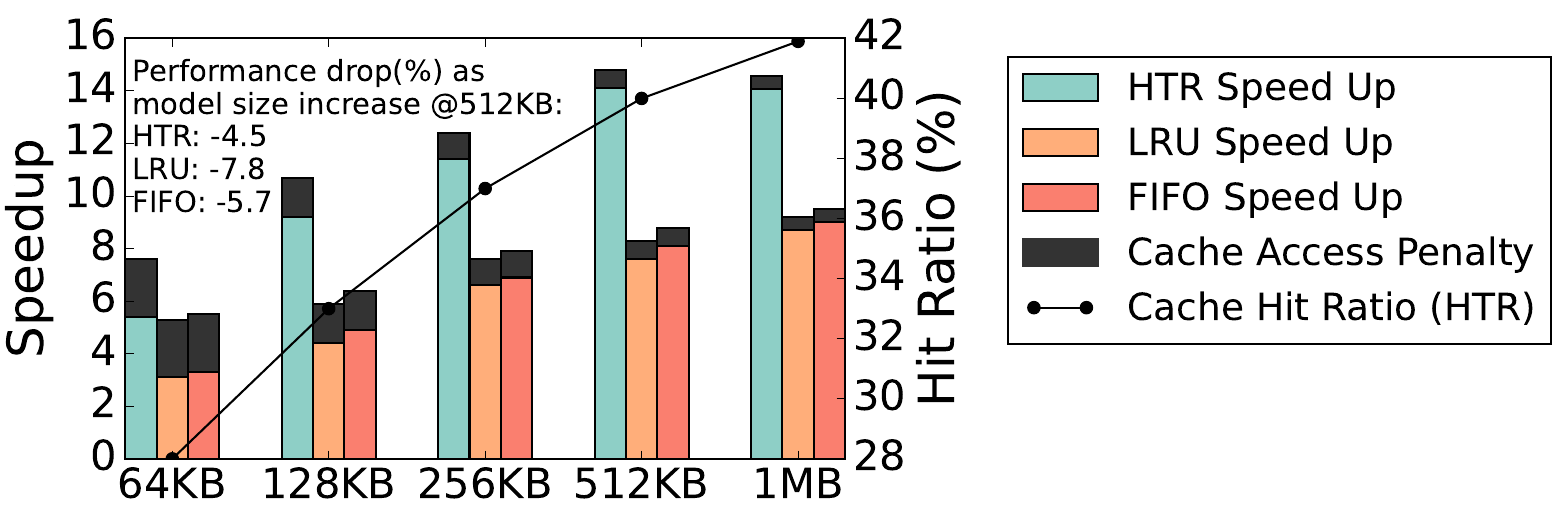}
    \caption{\textbf{Comparison of HTR performance across different cache sizes and replacement strategies.}}
    \label{fig:multifs}
\end{figure}

\subsubsection{On Switch Buffer Capacity}
\label{sec:capacityreslt}
As shown in Figure~\ref{fig:multifs}, the speedup numbers demonstrate how each caching strategy benefits system performance by reducing access times compared to the baseline (no caching). As the model size increase from RMC1 to RMC4, our HTR's maximum gain reduces from 19.3\% to 14.8\% with 512KB SRAM, due to larger footprint of the trace. In the largest model (RMC4), our hottest recording strategy (HTR) (\S \ref{sec:buffer}) generates better performance scaling with increasing cache size, achieving a speedup ranging from 7.6\% to 14.8\%, as the cache size increases from 64 KB to 512 KB. However, for the HTR strategy, a larger cache size (1 MB) results in performance degradation due to the absence of a significant increase in cache hit ratio (41.9\%) for the 1 MB cache, coupled with an increase in cache hit latency. This suggests that HTR with a cache size of 512KB effectively leverages data locality. In contrast, the LRU and FIFO strategies exhibit more modest improvements.

\begin{figure*}[t]
    \centering
    \begin{minipage}{0.4\textwidth}
        \centering
        \includegraphics[width=\linewidth]{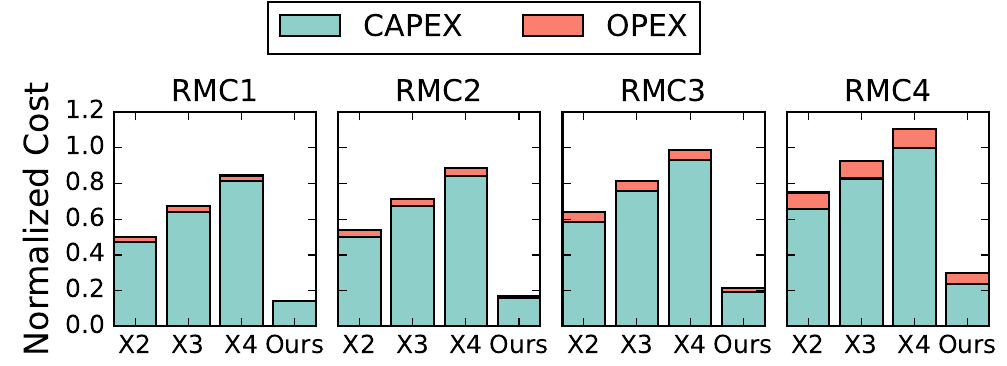} 
        \caption{\textbf{TCO under different models with increasing GPU budgets. e.g., X2 means 2 GPUs.}}
        \label{fig:tco}
    \end{minipage}\hfill
    \begin{minipage}{0.25\textwidth}
        \centering
        \includegraphics[width=\linewidth]{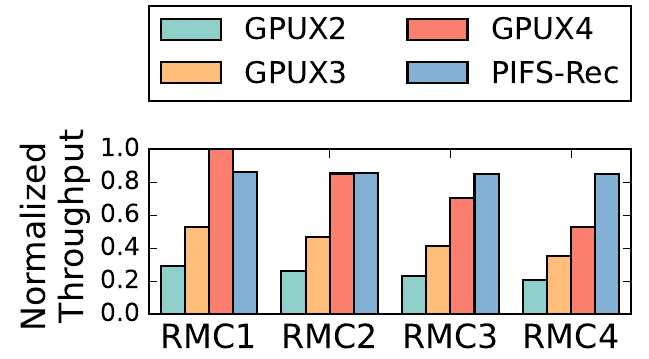}
        \caption{\textbf{Normalized throughput using different models.}}
        \label{fig:throughput}
    \end{minipage}\hfill
    \begin{minipage}{0.3\textwidth}
        \centering
        \scriptsize 
        \setlength{\tabcolsep}{1pt} 
        \begin{tabular}{@{}lcc@{}}
            \toprule
            System & Power/Area \\
            \midrule
            RecNMP-base(X8)~\cite{ke2020recnmp} & 75.4 mW / 215984 um² \\
            \midrule
            \textbf{PIFS-Rec Breakdown} & \\
            Process Core      & 9.3 mW / 33709 um² \\
            Control Logic + Registers     & 3.2 mW / 73114 um² \\
            \midrule
            On Switch Buffer & 15.2 mW / 2.38 mm² \\
            \bottomrule
        \end{tabular}
        \caption{\textbf{Hardware overheads.}}
        \label{tab:t3}
    \end{minipage}
\end{figure*}

\subsubsection{Page Management} 
\label{subsec:page_eval}
We collected the memory address ranges from the trace file and divided it into 4KB chunks as pages in the OS. In Figure~\ref{fig:plot4-8} (a), the embedding migration (\S \ref{sec:embeddingspreading}) strategy shows optimal performance at a 35\% migrate\_threshold, reducing latency by 14\% due to fewer page movements. Higher thresholds increase the embedding migration and degrade the performance (in fact, the migration cost increases from 1.67\% to around 10\% when we increase migrate\_threshold from 10\% to 50\% using page block). Our cache-line granular migration approach (\S \ref{sec:p2p}) outperforms standard OS page migration (page block) by up to 5.1$\times$, which decrease the migration cost to less than 2\%. We calculate the standard deviation (Std Dev) for access frequencies before and after the embedding migration (Figure~\ref{fig:plot4-8} (b)) to quantify the variability and assess the impact of 35\% migrate\_threshold. The standard deviation of access frequencies after the embedding migration drops from 20.6 to 7.8. This suggests that the PM effectively harmonizes access frequencies among the devices, leading to more uniform access frequency distributions and higher I/O parallelism. In Figure~\ref{fig:plot4-8} (d), for page swapping strategy (\S \ref{sec:em}), the best performance was observed with a cold\_age\_threshold of 16\% -- leading to a 12\% lower latency than TPP~\cite{maruf2023tpp}, which contributes to less page migration cost. The average migration cost decreases from around 8\% to 1\%. However, increasing the threshold further results in certain hot pages not being migrated to the local DRAM, consequently degrading the overall performance.

\begin{table}[htbp]
    \centering
    \caption{Hardware specifications.}
    \label{tab:t4}
    \scriptsize 
    \setlength{\tabcolsep}{1pt} 
    \begin{tabular}{llrr}
        \toprule
        \textbf{Hardware} & \textbf{Spec} & \textbf{TDP} & \textbf{Price} \\
        \midrule
        Server CPU~\cite{amd_epyc_9654} & AMD EPYC™ 9654 96C@2.4GHz & 360W & \$4,695 \\
        DIMM \& CXL mem~\cite{memverge_cxl_use_case} & per GB, DDR4 & 21.6W (64GB) & \$4.90 \\
        DIMM~\cite{memverge_cxl_use_case} & per GB, DDR5 & 24W (64GB) & \$11.25 \\
        NIC~\cite{juniper_qfx10002} & NVIDIA ConnectX-6@200Gbps IB & 23.6W & \$1,900 \\
        SWITCH~\cite{nvidia_connectx_6} & Juniper QFX10002-36Q @100Gbps & 360W & \$11,899 \\
        SWITCH + PUs~\cite{intel_tofino} & 3.2Tbps, 2 pipelines (ASIC) & 400W & \$13,039 \\
        GPU~\cite{nvidia_a100} & NVIDIA A100 80GB PCIe HBM2e & 300W & \$18,900 \\
        \bottomrule
    \end{tabular}
    
    \vspace{1ex} 
    \noindent\textit{Note:} The prices shown here are subject to market fluctuations and may not accurately reflect the actual procurement prices.
\end{table}

\subsection{Hardware Overheads}
We compare the power consumptions and hardware overheads with DLRM-dedicated system RecNMP, and the traditional DRAM based solutions. Since the previous work use different fabrication processes and EDA platforms that enabling different post-design optimization strategies, we keep the same functions as the prior work describe and map them to our fabrication process. We use Synopsys Design Compiler (DC) with a 1GHz clock to estimate the area and power consumption values. We also use this information to calculate the total energy consumption using conventional 45nm technology. To evaluate the energy consumption of standalone DIMMs with CPUs, we use Cacti-3DD~\cite{CACTI-3DD} for memory devices and Cacti-IO~\cite{Cacti-io} for the off-chip input/output operations at the DIMM level. 
In comparison to the prior solution that solely based on conventional DIMMs and CPU, PIFS-Rec reduces the energy consumption by 15.3\% on average. As shown in Figure~\ref{tab:t3}, PIFS-Rec reduces the power 2.7$\times$ compared to RecNMPs. PIFS-Rec requires 2.02$\times$ less area than an equivalent RecNMPs (x8) configuration with the same cache buffer.

\subsection{Cost and Performance Analysis}
We evaluate the performance of the parameter server over a simulation using one CPU-based server and a GPU server equipped with four A100 GPUs, obtaining power information using Nvidia-smi~\cite{Nvidia-SMI}. We conservatively estimate -- CXL memory's power consumption is 90\% of the local DRAM.

\textbf{TCO Model:} We assess capital expenditure (CAPEX) (Table \ref{tab:t4}) for RDMA hardware acquisition and operational expenditure (OPEX), including three years of power usage. Traditional setups involve a CPU in the GPU server along with NICs and a network switch. PIFS-Rec uses a CPU and fabric switch. We estimate the cost of a fabric switch considering the price of a standard network switch with an Intel Tofino core~\cite{intel_tofino}. We get power costs from the network’s standalone consumption and DC analysis data. Figure~\ref{fig:tco} demonstrates that PIFS-Rec offers superior TCO benefits compared to traditional GPU-based systems. For models with a few hundred parameters (RMC1), PIFS-Rec is 3.38$\times$ more cost-effective. Even for the largest models (RMC4) utilizing one GPU, PIFS-Rec is 2.53$\times$ cheaper. For instance, deploying RMC4 on a 2TB system with 64GB DIMMs requires \$27,769 to build a PIFS-Rec system, whereas a parameter server with a single GPU costs \$57,639. Assuming an energy cost of \$0.05 per-KWh~\cite{blsstrategies2023}, PIFS-Rec can save an additional \$2,332.14 in OPEX over three years. In a traditional GPU system, memory cost increases with the model size, whereas in our system, TCO benefit converges to the cost-benefit of DIMM and CXL memory. For smaller models (RMC1), GPU provides better throughput (Figure~\ref{fig:throughput}). However, with a large memory footprint and vector size, when memory bandwidth on the parameter server becomes the bottleneck throughput drops. In contrast, PIFS-Rec demonstrates high robustness compared to parameter server-based solutions and outperforms a 4-GPU cluster by 1.6$\times$. To understand the margin gain, we calculate performance-per-watt (PPW). As the model size increases, the PIFS-Rec's PPW improves from 1.22$\times$ to 1.61$\times$, compared to a 4-GPU conventional parameter server-based system.


\section{Concluding Remarks}
Deep Learning Recommendation Models (DLRMs) consume extensive datacenter cycles and struggle with bandwidth due to large embeddings and growing parameters. This study introduces a Process-in-Fabric Switch (PIFS) strategy to enhance DLRM efficiency in CXL-based systems. By optimizing {\em both} hardware and software for large-scale DLRM workloads, our system PIFS-Rec notably surpasses existing solutions, achieving up to 3.89$\times$ greater efficiency compared to current systems and doubling that of comparable architectures.

\section*{Acknowledgment}

We thank the anonymous reviewers for their useful feedback. This work was supported in part by gifts from AMD, INC and PRISM,
one of the seven centers in JUMP 2.0, a Semiconductor Research
Corporation (SRC) program sponsored by DARPA.

\bibliographystyle{./IEEEtran}
\bibliography{./IEEEabrv,./IEEEexample}

\end{document}